\documentstyle[preprint,revtex]{aps}  
 \def\bS{{\bf S}}
\def\bsigma{\sigma\kern-8pt\hbox{$\sigma$}\ }
\def\bdelta{\delta\kern-6pt\hbox{$\delta$}\ }
\def\bsupdelta{{^\delta\kern-6pt\hbox{$\delta$}}\ }
\def\bOmega{\Omega\kern-8pt\hbox{$\Omega$}\ }
\def\bell{\ell\kern-4pt\hbox{$\ell$}\ }

\def\br{{\bf r}}

\def\bk{{\bf k}}
\def\bq{{\bf q}}

\def\bR{{\bf R}}
\def\bi{{\bf i}}

\def\bl{{\bf l}}
\def\mbm{{\mbox{\boldmath $m$}}}
\def\mbJ{{\mbox{\boldmath $J$}}}
\def\mbp{{\mbox{\boldmath $p$}}}

\begin{document}
\draft
\begin{title}
HOLES IN A TWO--DIMENSIONAL\\
QUANTUM ANTIFERROMAGNET
\end{title}

\author{ Lu YU}
\begin{instit}
International Centre for Theoretical Physics, Trieste, Italy\\
and\\
Institute of Theoretical Physics, Chinese Academy of Sciences,\\
Beijing 100080,  China.
\end{instit}

\author{ Zhao-Bin SU}
\begin{instit}
Institute of Theoretical Physics, Chinese Academy of Sciences,\\
Beijing 100080, China
\end{instit}

\author{ Yan-Min LI}
\begin{instit}
Department of Physics, University of Warwick,\\
Coventry, CV4 7AL, United Kingdom
\end{instit}

\begin{center}
( To appear in Chinese Journal of Physics)
\end{center}

\begin{abstract}
A brief review is presented on the studies of the  hole motion in a
two--dimensional quantum antiferromagnet. An extended introduction is
given to cover the background of the problem. The quantum
Bogoliubov--de Gennes formalism which treats the local distortion of the
spin configuration and the quantum renormalization process on an equal
footing, is outlined. The latest development on the central issue, whether a
hole can freely propagate on an antiferromagnetic background, is
overviewed.
\end{abstract}
\newpage

\section{INTRODUCTION}

The discovery of high temperature oxide superconductors \cite{1} has
stimulated great interest in the studies of the interplay between
magnetism and superconductivity. It has been established now that the
reference compounds like $La_2 Cu O_4$ in the case of lanthanum family
or $YBa_2Cu_3O_6$ for the yttrium family are antiferromagnetic (AF)
insulators \cite{2}. Upon doping the AF long range order fades away and
superconductivity occurs below certain critical temperature. A generic
phase diagram is shown in Fig.1. Whether the superconductivity is caused by
magnetic fluctuations is a separate issue. It is, however, crucial to
understand the behaviour of charge carriers on an AF background.

This is not a trivial problem, because
according to the single electron
theory, the reference compound has a half--filled valence band, and hence
should be a metal \cite{3}, whereas it is indeed a
Mott insulator. Unlike the simple band insulators, the energy gap for charge
excitations in this type of insulators originates from strong electron
correlations. In the early days of the high--$T_c$ heat wave Anderson
suggested that a doped Mott insulator described by the one--band Hubbard
model should contain the basic physics of the high--$T_c$ superconductors
\cite{4}. This model, independently proposed in 1963 by Gutzwiller, Hubbard
and Kanamori \cite{5}, to describe the competition between delocalization
and correlations in narrow--band transition metals has a simple
looking:
\begin{equation}
H=-t\sum_{\langle nn \rangle,\sigma}\ C^\dagger_{i\sigma}\ C_{j\sigma}
+U\ \sum_i\
C^\dagger_{i\uparrow}\ C_{i\uparrow}\ C^\dagger_{i\downarrow}\
C_{i\downarrow},
\end{equation}
where $\langle nn \rangle$ indicates summation
over nearest neighbours, the indices  $i, j$ run
over $N$ lattice sites, $C^\dagger_{i\sigma}(C_{i\sigma})$ are the electron
 creation (annihilation) operators of spin $\sigma$ at site $i$, and
the two--body interaction is on--site and repulsive $(U> 0)$.
There is a vast literature devoted to the study of this model in the last 30
years, especially after 1987 \cite{6}.

In the large $U$ limit, it is more convenient to use a canonical
transformation to project out the doubly occupied sites costing energy
$U$ and to obtain an equivalent, the so--called $t-J$ model \cite{7}:
\begin{equation}
H=-t\sum_{\langle i,j\rangle ,\sigma}(C^\dagger_{i\sigma}\ C_{j\sigma} +
h.c.)+J\sum_{\langle i,j\rangle}\left(\bS_i\  \cdot \bS_j-{1\over 4}\ n_i\
n_j\right),
\end{equation}
where restriction to the Hilbert space without double occupancy is assumed
and $n_i=\sum\limits_\sigma\ n_{i\sigma}$ is the density operator at site
$i$ with $n_{i\sigma}=C^\dagger_{i\sigma} C_{i\sigma}$. The summation
in the second term is carried out only over nonrepeated bonds $\langle
ij\rangle$ and under this convention $J=4t^2/U$. In deriving (2) from (1) a
three--site term has been neglected which should not modify substantially
the physics, especially when the deviation from half filling $\delta$ is
small. The spin density operator $\bS_i$ is defined as:
$\bS_i=C^{\dagger}_i\ {\displaystyle{\bsigma\over 2}}\ C_{i}$, where
${\bsigma}$ are the Pauli matrices. At half--filling (one electron per site)
the hopping term is absent, so the $t-J$ model (2) becomes the well--known
Heisenberg model for spin 1/2 in the fermion representation up to a
constant. Of course, the $t-J$ model (2)
can be studied independently, even in the limit $t\ll J$, but the
correspondence with the Hubbard model is established only in the limit
$t\gg J$, or $U\gg t$. In this article we  consider a slightly generalized
$t-J$ model with anisotropic exchange interactions, {\it i.e.},
\begin{equation}
H=-t\sum_{\langle i,j\rangle ,\sigma}(C^\dagger_{i\sigma}\
C_{j\sigma}+h.c.)+J\sum_{\langle i,j\rangle}\left[ S^z_i\ S^z_j+\alpha
(S^x_i\ S^x_j+S^y_i\ S^y_j)\right],
\end{equation}
where $\alpha =0$ corresponds to the Ising limit while $\alpha =1$ is the
Heisenberg case.

It is now well--established that the essential ingredient in cuprate
superconductors is the two--dimensional $CuO_2$ plane \cite{8}. The
$Cu-O$ two band model, proposed by several authors \cite{9} would be more
appropriate for the description of physical properties of this plane.
However, as shown by Zhang and Rice \cite{10}, the low--energy physics is
determined by the singlet state formed by the additional hole on oxygen
with the existing hole on copper. The hopping of this singlet is described by
an effective $t-J$ model which makes this model  even more
pertinent to high--$T_c$ superconductors. Of course, there are aspects of
the two--band model, like the formation of charge--transfer polarons
\cite{11}, which cannot be described by the one--band $t-J$ model.

The standard BCS theory of superconductivity is based on the assumption
that the Landau Fermi liquid theory can adequately describe the normal
state of superconductors \cite{12}. According to the Fermi liquid theory
\cite{13} the elementary excitations of an interacting fermion system can
be mapped onto a noninteracting Fermi gas. These quasi-particles
have
spin 1/2, charge $e$ and are well defined,  which means that the real part
of the self-energy (of the
order temperature $T$ or external frequency $\omega$ ) is much greater than the
imaginary part (proportional to the square of the corresponding small
quantities).

The propagator of these quasi--particles has a nonvanishing residue $Z$ at
the pole $\omega_k$, $i.e.$, the spectral density can be written as
\begin{equation}
A(\bk ,\omega )=Z_k\delta (\omega -\omega_k)+{\rm
incoherent\ part},
\end{equation}
which means existence of a well--defined Fermi surface where the momentum
distribution has a finite jump equal to $Z_{k_F}$. This was first observed by
Migdal \cite{14} and later elaborated by Luttinger \cite{15} who also
showed that the volume enclosed by this Fermi surface should be the same
as for noninteracting systems.

It is not clear at all whether  charge carriers on an AF background will
share all these properties, because the starting AF insulating state exhibits
strong electron correlations. The central issue is whether a single hole
 can freely propagate on an AF background, $i.e.$, whether $Z$ is
different from zero or
not. Of course, there is a finite concentration of charge carriers in real
superconductors, but we will limit ourselves in this article to the
single hole issue with the hope that the solution of this simple looking
problem will shed some light on the underlying physics of high--$T_c$
superconductors. The reader is referred to the relevant literature \cite{8}
for issues of finite doping concentration.

In this brief review we will summarize our studies on the single hole
problem with a rather detailed introduction  and
an overview of the recent developments. The discussion will be developed
around the central issue mentioned above. The rest of the paper is organized
as follows. Section II is devoted to the background material, while various
approaches to the single hole problem are discussed in Section III. The
quantum Bogoliubov--de Gennes (BdeG) approach developed by us to treat
the distortion of the spin background and quantum fluctuations on an equal
footing, is summarized in Section IV. The latest developments and
perspectives are discussed in the final section.

\section{BACKGROUND}
\noindent 2.1 The reference state

The ground state of the large $U$ Hubbard model at half--filling or its
equivalent -- the Heisenberg model, depends strongly on the dimensionality.
In one dimension (1D) an exact solution for the spin 1/2 Heisenberg chain
was obtained by Bethe \cite{16} using the famous Bethe Ansatz (a term
coined by C.N. Yang and C.P. Yang \cite{17}). An exact solution for the
Hubbard model at arbitrary filling was obtained by Lieb and Wu \cite{18},
using the same ansatz. The ground state at half--filling is an insulating
state for arbitrary $U$ without AF long range order, while the spin--spin
correlation function shows a power law decay.

There are no exact solutions beyond 1D. However, some rigorous results are
available. In particular, Marshall showed in 1955 \cite{19} that the ground
state for $S=1/2$ systems on a bipartite lattice (2--sublattice system with
all nearest neighbours  of a given site
belonging to the second sublattice) should be a singlet
state. This is known as the ``Marshall sign
rule''. It has been proved  recently by Kennedy, Lieb and Shastry
\cite{20} that there is AF long range order in three dimensions (3D) for
$S=1/2$ system at $T=0$, which vanishes at certain critical temperature. For
two dimensions (2D) the AF long range order in the ground state has been
proved for a square lattice in the case $S\geq 1$ or $S=1/2$, but $0\leq
1/\alpha\leq 0.13$ and $1/\alpha > 1.78$ \cite{21}. For 2D $S=1/2$
Heisenberg model, relevant for the high--$T_c$ problem, neither existence,
nor nonexistence of AF long range order has been proved. Nevertheless,
there is strong evidence from exact diagonalization of small clusters and
quantum Monte Carlo simulations that such order does exist for a 2D square
lattice at $T=0$ \cite{22}. However, the staggered magnetization is reduced
to $m\approx 0.31\pm 0.02$ due to quantum fluctuations. We will take this
as the starting point of our discussion. Of course, there cannot be any long
range order at finite temperatures for the 2D Heisenberg model, as follows
from the Mermin--Wagner theorem \cite{23}.

Historically, the question of the ground state for spin 1/2 systems has been
under active discussion. Since such antiferromagnets were not found for
a long time, Landau and Pomeranchuk suspected that there might be
no AF long range
order  and the elementary excitations might be spin 1/2
fermions for $S=1/2$ AF systems \cite{24}. Anderson \cite{25} proposed in
1973, the so--called Resonant--Valence--Bond (RVB) state as a possible
candidate for the ground state  of such systems, most probably
for a frustrated lattice. The RVB state
is a condensate of singlet pairs without AF long range order. In 1987 he
reformulated his proposal in terms of a BCS--like variational wave function
\cite{4}. Since then a large amount of literature
has been devoted to this issue
\cite{6,8}. The general consensus now is:
There is an AF long range order in the ground state of 2D spin 1/2
Heisenberg model, at least for a square lattice.
However, the generalized RVB state with long range pairs
and expressed in terms of spin variables instead of fermions
to satisfy the Marshall sign
rule \cite{19}, has a very close energy and  similar
correlation functions as the
AF state \cite{26}.
\vskip0.5truecm

\noindent 2.2 Spin wave theory

It is well known that the classical and the quantum mechanical description
for a ferromagnetic ground state is essentially the same because the order
parameter -- the spontaneous magnetization -- commutes with the
Hamiltonian. This is not true for the AF ground state, where
the order parameter -- the staggered magnetization -- does not commute
with the Hamiltonian, serving as a typical case of non trivial symmetry
breaking. The classical ground state would be a N\'eel state with
alternating spin alignment, which is true for the Ising model, but not for the
 Heisenberg case. For the latter one should take into account the
quantum fluctuations (or spin zero--point motion).

The spin operators satisfy the following commutation relations
\begin{equation}
\left[ S^\alpha_k , S^\beta_\ell\right] =i\
\varepsilon^{\alpha\beta\gamma}\ S^\gamma_\ell\ \delta_{k\ell},
\end{equation}
where $\alpha ,\beta ,\gamma = x,y,z$, \  $\varepsilon^{\alpha
\beta\gamma}$ is the unit antisymmetric tensor. If we introduce the spin
raising and spin lowering
operators
$S^+_k =S^x_k +iS^y_k , S^-_k=S^x_k -iS^y_k$, for
$S=1/2$ they satisfy the commutation relations
\begin{equation} \left[ S^+_k ,S^-_k\right]_+ =1,
\end{equation}
\begin{equation}
\left[ S^+_k,S^-_\ell\right] =0,\quad k\not=\ell
\end{equation}
i.e., they behave like fermions on the same site, and as bosons on different
sites. These are {\em hard--core bosons}, or Pauli operators. We will come
back to this special property of the spin 1/2 operator at the end of this
subsection, but now we  consider the case of arbitrary $S$.

Holstein and Primakoff \cite{27} have proposed a nonlinear transformation
to convert the spin operators into bosons
\begin{equation}
S^z=S-b^\dagger b,\quad S^-=\sqrt{2S}\ b^\dagger (1-b^\dagger b/2S)^{1/2},
\end{equation}
and one can then expand  them in terms of $b^\dagger b/2S$.
This is the so-called linear spin wave
theory, if only the leading term is considered. Apparently, the expansion
parameter here is $1/2S$ which is 1 for  $S=1/2$. However, the actual
expansion parameter turns out to be $1/Sz$ where $z$ is the coordination
number \cite{22}. With this understanding, the application of the linear
spin wave theory to the 2D $S=1/2$ Heisenberg model is more or less
justified.

Anderson and Kubo \cite{28} have applied this theory to the
AF case. Since its presentation is available in standard textbooks
\cite{29}, we will outline here only the main results for further reference.
Using the Holstein--Primakoff transformation
\begin{eqnarray}
S^+_i &=& \sqrt{2S}\ b_i,\quad S^z_i =S- b^\dagger_i\ b_i,\quad
i\in\uparrow\ {\rm sublattice} \nonumber \\
&& \\
S^+_j &=& \sqrt{2S}\ b^\dagger_j,\quad S^z_j=-S+ b^\dagger_j\
b_j,\quad j\in\downarrow\ {\rm sublattice}, \nonumber
\end{eqnarray}
the anisotropic Heisenberg Hamiltonian becomes
\begin{equation}
H_J={J S}\ \sum_{\langle i,j\rangle} (2\ b^\dagger_i\ b_i+\alpha
(b_i\ b_j+b^\dagger_i\ b^\dagger_j)) -{NJz\over 4}
\end{equation}
which, in turn, can be diagonalized using the Bogoliubov transformation
\begin{equation}
b_k =u_k\ \beta_k + v_k\ \beta^\dagger_{-k}
\quad {\rm with}\quad \vert u^2_k\vert -\vert v_k\vert^2=1
\end{equation}
to obtain
\begin{equation}
H_J=\sum_k  \omega_k \left(\beta^\dagger_k\ \beta_k+{1\over 2}\right)
-{3NJz\over 4}, \end{equation}
where
\begin{equation}
\omega_k ={JSz}\ (1-\alpha^2\gamma^2_k)^{1/2},\quad
\gamma_k ={1\over z}\ \sum_\delta\ e^{i k \cdot \delta}
\end{equation}
with $\bdelta$ as the nearest neighbour lattice vector. The spin--wave
vacuum $\vert 0\rangle$ can be expressed in terms of the classical N\'eel
vacuum $\vert N\rangle$ as
\begin{equation}
\vert 0\rangle =\prod\limits_k\ u^{-1}_k\exp\left(
-\lambda_k\ b^\dagger_k\ b^\dagger_{-k}\right)\ \vert
N\rangle \end{equation}
with
\begin{eqnarray}
\lambda_k & = & -v_k /u_k, \nonumber \\
u_k & = &\left\{ {1\over 2}\left[ (1-\alpha^2\gamma^2_k
)^{-1/2}+1\right]\right\}^{1/2}, \nonumber \\
v_k & = & -sgn (\gamma_k )\left\{ {1\over 2}\left[
(1-\alpha^2\gamma^2_k )^{-1/2}-1\right]\right\}^{1/2}.
\end{eqnarray}

The spin--wave spectrum for the Heisenberg case $\alpha =1$ is linear in
$k$
\begin{equation}
\omega_k =v\vert\bk\vert
\end{equation}
with the spin--wave velocity $v=2\sqrt d\ JS$. This is the Goldstone mode
originating from the breaking of O(3) continuous symmetry. One can easily
calculate the reduction of the order parameter due to the quantum
fluctuations
\[ \langle S^z\rangle \ = S-\langle b^\dagger_i\ b_i\rangle \]
and the correction is approximately 0.1 for a 3D cubic lattice,
0.2 for a 2D square lattice and
diverges for 1D.

There is one serious difficulty with the spin wave theory, namely, the
hard--core nature of the spin operator, or equivalently, the single
occupancy constraint is not well preserved. In fact, one can calculate the
expectation value \cite{30}
\[
F\equiv\vert\langle [ S^+_i, S^-_i]_+ \rangle -1\vert
=2\vert\langle [ b^\dagger_i\ b_i-(b^\dagger_i\
b_i)^2]\rangle\vert  \approx 1.06 \]
as  temperature  $T\to 0 $,
which should vanish if the constraint is strictly obeyed.

There have been many attempts to improve the linear spin wave theory.
Dyson and Maleev \cite{31} have introduced another transformation of spin
operators into boson operators to consider the kinematic constraints.
Arovas and Auerbach have used the Schwinger boson transformation to
represent the spin operators and the path integral technique  to
impose the constraint \cite{32}. Takahashi \cite{33} has applied the
Dyson--Maleev transformation to quantum AF to impose the kinematic
constraints. His results turned out to be essentially the same as those of the
Schwinger boson approach. As a common weakness of these treatments,
the single occupancy constraint is imposed only on average. An interesting
attempt was made by Wang \cite{34} who used the Jordan--Wigner
transformation generalized by several authors \cite{35} to 2D to represent
the hard--core boson in terms of spinless fermions. In this representation
the hard--core constraint is satisfied exactly and very good results can be
obtained even in the mean field approximation. Recently, this
transformation has been extended to the case away from half--filling and a
number of interesting results have been obtained \cite{36}.
\vskip0.5truecm

\noindent  2.3 Nagoaka theorem and ferromagnetic polaron

The phase diagram of the Hubbard model or the $t-J$ model upon doping is a
complicated problem. However, there is one rigorous theorem proven by
Nagaoka \cite{37} which can be stated as follows: For a bipartite lattice
with one hole in the infinite $U$ limit, the ferromagnetic state with
maximum spin is the ground state.

Instead of presenting its proof we give here some simple argument in
favor of it.
The main difficulty of the Hubbard model in the infinite $U$ limit is the
enormous spin degeneracy at exact half--filling. The presence of a single
hole can lift the degeneracy. Intuitively, it is clear that the hole can gain
maximum kinetic energy if all spins are aligned ferromagnetically. This
maximum energy is $zt$ with $z$ as the coordination number. Using a set of
states with spins permuted by single hole hopping around a closed loop,
Nagaoka could prove this statement rigorously for a bipartite lattice where
all nearest neighbours of a given site belong to the second sublattice and
the sign of $t$ is irrelevant.

This theorem has nontrivial consequences. First of all it is not true for 1D
where no closed loops are available. Secondly, a generalization to finite
hole concentration  in the thermodynamic limit seems to be natural, but has
confounded all analytical attempts. The Pauli exclusion principle is
irrelevant for a single hole, but it makes the problem  messy even for two
holes. The result depends strongly on the boundary conditions. For the periodic
boundary condition a ``spiral phase'' is preferred, whereas the
ferromagnetic state survives for ``averaged'' boundary conditions \cite{38}.
Numerically, there is strong evidence that a ferromagnetic state with
non--saturated magnetization is the ground state at small, but finite
doping concentration as $U\to\infty$ \cite{39}.

In real systems $U$ is finite and $J$ is different from zero. A more
pertinent question would be: How big is the ferromagnetic region around a
hole as a result of balancing between the kinetic energy\  $t$\ favoring the
ferromagnetic configuration and the exchange energy $J$ favoring the AF
alignment of neighbouring spins. A rough estimate for the 2D Ising square
lattice gives the hole energy as \cite{40}
\begin{equation}
\varepsilon_h/t=-4+6.03(J/t)^{1/2},
\end{equation}
where the radius of the ``ferromagnetic'' disc diverges as $(J/t)^{-1/2}$.
For a small cluster $(4\times 4)$, the ferromagnetic state with maximum
spin is materialized only for $J/t<0.075$ \cite{41}. We will call such a
formation  ``ferromagnetic polaron''.

In view of the Nagaoka theorem, the problem of single hole motion on an
``AF'' background is irrelevant in the limit $U=\infty$, or $J=0$. However,
apart from very small $J/t$, when the energy of ferromagnetic polaron (17)
is lower than the competing configuration of spin polaron (to be considered
later), the single hole problem is an important issue and will  be dealt
with in the rest of this article.
\vskip0.5truecm

\noindent  2.4  One--dimensional case

Although the exact solution of the 1D Hubbard model was obtained by Lieb
and Wu 25 years ago \cite{18}, much of the rich physics contained in this
solution has been appreciated only during the recent intensive studies on
the high--$T_c$ problem. These results  have no direct relevance for the
single hole issue in 2D, but they are very instructive and inspiring by
providing a profound example with a number of
peculiar properties, in many ways
contrary to our intuition. Therefore, we outline here some of these
properties, which were emphasized by Anderson \cite{42}.

One of the distinguished features is the absence of quasi--particles in the
standard sense. Using the Bethe Ansatz solution \cite{18}, one can
calculate the wave function renormalization $Z$, defined as  square of
the overlap integral between the $N$--particle ground state wave function
plus one free particle in state $k$ with the $N+1$--particle ground state
wave function, {\it i.e.} \cite{42}
\begin{equation}
Z_k =\vert\langle\psi_{k\sigma}(N+1)\vert
C^\dagger_{k\sigma}\ \psi_G (N)\rangle\vert^2\sim e^{-{1\over
2}({\delta\over\pi})^2{\rm ln}\ N},
\end{equation}
where the phase shift $\delta ={\pi\over 2}$ in the large $U$ limit. In the
thermodynamic limit, $N\to\infty ,Z_k\to 0$, hence the quasi--particle does
not exist as such. This is another example of the ``infrared catastrophe'',
suggested by Anderson earlier \cite{43}.

The main physical origin for a vanishing  $Z$ is the spin--charge
separation which can be explicitly seen in the asymptotic form of the Bethe
Ansatz wavefunction in the large $U$ limit, derived by Ogata and Shiba
\cite{44}
\begin{equation}
\psi (x_1,\dots x_N)=(-1)^Q\ \det [\exp (ik_j x_{Q_j})]\ \phi (y_1,\dots y_M),
\end{equation}
where $x_{Q_j}$ are sites of spinless fermions (holons), while $y_M$ are
the down--spin sites of a ``squeezed'' Heisenberg chain with all hole sites
being removed. When a hole is added at $k_F$, it decays immediately into a
holon (charge $e$ without spin at $2k_F$) and a spinon (spin 1/2 and no
charge). The vanishing of $Z$ reflects the absence of charge for excitations
at $k_F$.

Moreover, one can calculate various correlation functions and the
single--particle Green's function \cite{44,45,42}. All of them show a power
law decay with exponents depending on the interaction. It is remarkable
that in spite of the vanishing of $Z$, hence the jump of the momentum
distribution at the Fermi surface, there is still a singularity of the latter
at $k_F$ (not $2k_F$), {\it i.e.},
\begin{equation}n_k ={1\over 2}-const \vert k-k_F\vert^\alpha\
sgn(k-k_F),
\end{equation}
where the exponent $\alpha = 1/8$ for $U\to\infty$.

It turns out that the above described properties ($Z=0$, spin--charge
separation, power decay of the correlation functions, {\it etc.}) are rather
generic for 1D interacting systems. Haldane \cite{46} first realized the
existence of this large universality class and has coined a term ``Luttinger
Liquids'' to contrast them with the Landau theory of Fermi liquids.

Anderson suggested \cite{42} that the 2D Hubbard model should share these
properties and the experimental data on high--$T_c$ superconductors,
especially the normal state properties seem to be consistent with  this
assumption. Of course, the question is still widely open.
Clearly, the single hole problem cannot replace the issue of the
charge carrier behaviour at finite doping concentration, but it is a pertinent
question to consider in order to clarify the situation.

\section{VARIOUS APPROACHES FOR STUDYING THE SINGLE HOLE PROBLEM}

The question of charge carrier motion on an AF background and the possible
formation of self--trapped states have been discussed in the literature
long time back in connection with magnetic semiconductors \cite{47,48}. In
particular, de Gennes \cite{48} has considered the formation of a small
ferromagnetic region around the charge carrier which can propagate in an
AF under certain conditions. There were extensive studies on this
issue in Russia \cite{49,50},  summarized later in a review
article and a book \cite{51}. Unfortunately, these research works are not
well known in the Western literature. We will, therefore, mainly discuss
those studies which have shaped our current understanding of the problem.
\vskip0.5truecm

\noindent 3.1 Strong coupled Ising case (the Brinkman--Rice limit)

Consider first the Ising case. The hopping of a hole on a N\'eel background
would create a string of ``wrongly'' aligned spins (see Fig.2) costing energy
of the order $\ell J$, where $\ell$ is the length of this string. To avoid this
energy expense, Brinkman and Rice \cite{52} have introduced the so--called
``retraceable path approximation'' under which the hole hops back exactly
the same way as it left the starting point, so there is no difference in the
initial and final spin configurations. In the atomic limit of the Hubbard
model $(U\to \infty$ or $J=0$), they have summed up all these retraceable
paths (without closed loops) and have come to the following conclusions:

\noindent 1) The hole cannot propagate freely, i.e., $Z=0$, and the Green's
function is $k$--independent, i.e.
\begin{equation}
G(\omega )=\omega^{-1}\left\{ 1-{z\over (z-1)}\left[{1\over
2}-\left({1\over 4}-{(z-1)t^2\over\omega^2}\right)^{1/2}\right]\right\}^{-1}
\end{equation}
with $z$ as the coordination number and $t$ as the hopping integral.

\noindent 2) The density of states $\rho (\omega )$ has a square--root
singularity with  a 25\% narrowing compared with the original bare
bandwidth, namely,
\begin{equation}
zt\ \rho (u)={1\over\pi}\left[ (5-9u^2)^{1/2}/(1-u^2)\right] ,\quad
u=\omega /zt,
\end{equation}

\noindent 3) The electric conduction is due to diffusive motion of holes and
the mobility can be calculated.

In spite of the great success of the Brinkman--Rice approach, there are two
serious limitations.

\noindent i) They have ignored the closed loops which turned out
to be very important. As noted by Trugman \cite{55}, a hole can travel
around a square one and half times without disturbing the spin background
and finds itself translated to the next--nearest--neighbour (see
Fig. 3). This means
that the hole can ``unwind'' the string and self--generate a
next--nearest--neighbour hopping. Therefore, the full localization of
charge carriers in the Brinkman--Rice limit is an artifact of the
approximation.

\noindent ii) Their consideration for the spin dynamics is purely classical,
i.e. the emission and reabsorption of spin excitations are not allowed. As
we will see later, the renormalization effect of the hole motion due to
these processes is essential even for the Ising case.

If $J/t$ is finite, but small, the ``string'' picture holds, and one can
calculate the bound states for the hole in this confining potential which
leads to an Airy equation \cite{50,53}. The semi-classical
ground state energy in 2D is then
given by
\begin{equation}
E_g/t=-2\sqrt 3 + 2.74(J/t)^{2/3},
\end{equation}
where the ``band narrowing'' effect ($2\sqrt3$ instead of 4) of Brinkman
and Rice has been taken into account. The energy difference between higher
levels is also $\sim (J/t)^{2/3}$. This string picture has been explored
further by many authors to interpret various physical properties \cite{54}.
Vollhardt {\em et al.} \cite{54a} have found that the retraceable path
approximation is exact in infinite dimensions and they have generated
a systematic $1/d$ expansion.

\vskip0.5truecm

\noindent 3.2 Quasi--classical approach (Nonlinear $\sigma$--model)

In Section 2.2 we outlined the traditional spin wave theory for AF. An
alternative way of describing the long--wavelength spin fluctuations for
large $S$ is to use the nonlinear $\sigma$--model, assuming the existence
of AF long range order. The order parameter--staggered magnetization
$\bOmega$ is a vector on a unit sphere, whereas the magnetization $\mbm$ is
a small quantity. In the long wavelength limit  (at the scale larger than
the lattice constant) the fluctuations are described by the nonlinear
$\sigma$--model \cite{56}:
\begin{equation}
H_{NL\sigma}={1\over 2}\int d^2r\left[\chi^{-1}\ \mbm^2+\rho
(\nabla\cdot\bOmega )^2\right] ,
\end{equation}
where $\chi$ is the magnetic susceptibility, while $\rho$ is the spin
stiffness. Their ratio $\rho /\chi =c^2$ is the square of the
spin--wave velocity. $\bOmega$ and $\mbm$ satisfy the commutation
relations
\[
[m^\alpha (x), m^\beta (y)]   =   i\ \varepsilon^{\alpha\beta\gamma}\
m^\gamma (x)\ \delta (x-y), \]
\begin{equation}
[m^\alpha (x), \Omega^\beta (y)]   =    i\
\varepsilon^{\alpha\beta\gamma}\ \Omega^\gamma (x)\ \delta (x-y)
\end{equation}
and the constraints
\begin{equation}
\bOmega^2=1,\qquad \mbm (\br )\cdot\bOmega (\br )=0.
\end{equation}
The spin current is defined as
\begin{equation}
\mbJ_a=\bOmega\times\partial_a\ \bOmega ,
\end{equation}
satisfying the equation of motion $\partial\mbm /\partial t=-\rho\
\partial_a\mbJ_a$.
This current is a vector in both spin and coordinate space.

Shraiman and Siggia \cite{57} have considered the coupling of the charge
carrier to the spin configuration in the quasi-classical
approximation $t\ll J$.
The spin background is assumed to be classical but the distortion of this
background due to the hole motion has been taken into account. In analogy
with the backflow in helium they considered the dipole moment generated
by the charge carrier
\begin{equation}
\mbp_a (\bq ) = \sum_k\sin k_a\ \psi^\dagger_{\bk-\bq /2}\
\bsigma\psi_{\bk +\bq /2},
\end{equation}
where $\psi_{\bk}$ is the spinor wave function of the hole, $\bsigma$ the
Pauli matrices. The $k$--summation is carried over the minimum energy
states. (Cluster calculations, as discussed in Section 3.4, show they are at
$\big(\pm {\pi\over
2},\pm{\pi\over 2}\big)$). The coupling to the spin current is written as
\begin{equation}
-g\sum_{a,\bq}\ \mbp_a(\bq )\cdot \mbJ_q(\bq ).
\end{equation}
They found that in the continuum limit
\begin{equation}
\bOmega\times\partial_a\ \bOmega =(\delta_{ab}-2\hat r_a\ \hat r_b)\
{\mbp_b\over r^2},
\end{equation}
which gives rise to a dipolar spin configuration (Fig.4). Based on this
result, they proposed that the commensurate AF phase is unstable with
respect to an incommensurate spiral phase upon doping. The pitch of the
spiral should be inversely proportional to the doping concentration
\cite{57}. Most of their results were derived under the assumption $J\gg t$,
but they have also used the Brinkman--Rice ``string'' picture to argue that a
similar conclusion should also hold in the opposite limit $J\ll t$, since
the symmetry properties are the major factor.

This approach also  has the  limitation
that the quantum fluctuations of spins have not
been included,  as  in the
Brinkman--Rice method.  As we will see in
the next subsection, inclusion of these
fluctuations will modify some conclusions in a substantial way. Also, the
spin distortion discussed by them is related to areas far away from the hole
location, {\it i.e.} it is a  long distance behavior.
As shown in the next section, the near--field distortion is quite
different.

This approach has also been used by several other groups to consider the
single hole problem \cite{58} with similar results.
In particular, a self--trapped
polaron with local ferromagnetic distortion is found when the hole state is
at the Brillouin zone center. The structure of polaron is much more
complicated if the hole state is at the zone boundary.

A semiclassical large $S$ expansion has also been employed by Auerbach
and Larson to study the single hole problem \cite{59}. Using the spin--hole
coherent--state path integral, they have generated a large $S$ expansion of
the $t-J$ model on a square lattice. The single--hole classical energy
is minimized by a small polaron solution.
In the parameter range $1<t/2JS<4.1$, the
polaron involves only five sites with one flipped spin at the center. The
hole density is 1/2 and 1/8 at the center and the nearest neighbour,
respectively. Contrary to Shraiman and Siggia, they have not found any
long--distance distortion of the AF background. They have also found that
the dynamical nearest neighbour hopping is suppressed and have recovered
an effective polaron model with  next--nearest--neighbour hopping only
\cite{60}.
\vskip0.5truecm

\noindent 3.3 Self--consistent Born approximation

There are two effects which have not been properly considered in the above
described approaches, namely, the quantum fluctuations of the Heisenberg
spin and the renormalization of hole motion due to interactions with spin
waves. These effects have been taken into account in the self--consistent
Born approximation carried out by Schmitt--Rink, Varma and Ruckenstein (SVR,
\cite{61}) and Kane, Lee and Read (KLR, \cite{62}).

The hopping of holes leads to disturbance of the spin background. Suppose
there is a hole on spin--up sublattice site. After hopping to spin--down
sublattice site, a down--spin appears on a spin--up lattice site,
corresponding to a local spin excitation. Likewise, a hole on spin--down
sublattice would request the neighbouring up--spin to flip first in order to
avoid frustration. Introducing hole operator $h_i$ to replace
$C^\dagger_{i\uparrow}$ on the spin--up sublattice, while representing the
corresponding operator $C^\dagger_{i\downarrow}=h_i\ S^-_i$ as a
composite operator (similarly for the spin--down sublattice), and then
performing the Holstein--Primakoff transformation, one obtains the
hopping Hamiltonian \cite{61,62}
\begin{equation}
H_t=t\sum_{\langle i,j\rangle} h^\dagger_ih_j\ (b^\dagger_j+b_i)\
\end{equation}
in the linear spin--wave approximation. A well--argumented derivation of
(31) has been given in Ref.64. The form of (31) is similar to the
electron--phonon interaction. It is natural then to compare the
spin--polaron under discussion  with the well--studied lattice polaron
\cite{64}.  However, there are important differences between
these two, as we will discuss in Section 4.2.

Comparing the Heisenberg Hamiltonian (10) and the hopping term (31),
one finds two different limits. When $J\gg t$, one can use the perturbation
theory, but this is a less interesting case. The alternative limit $J\ll t$,
corresponding to the large $U$ Hubbard model, cannot be treated
perturbatively. One has to perform at least a partial resummation to obtain
a meaningful result. This was done by SVR \cite{61} and KLR \cite{62}. They
have ignored the Hartree diagram, but have taken into account the Fock
term in the self--energy (Fig.5). Neglecting the crossing diagrams (Fig.6),
they can solve the corresponding Dyson equation. The self--energy part can
be written as
\begin{equation}
\Sigma (\bk ,\omega )={t^2z^2\over N}\ \sum_{\bk}\ (u_{\bq}\
\gamma_{\bk-\bq}+v_{\bq}\ \gamma_{\bk})^2\ G(\bk -\bq ,\omega
-\omega_{\bq}
 ),
\end{equation}
where the renormalized hole propagator
\begin{equation}
G(\bk ,\omega )=(\omega -\Sigma (\bk ,\omega )+i\ \delta )^{-1},
\end{equation}
while the definitions for $u_k,v_k,\omega_k,\gamma_k$ were given in
(13)--(15). One can then easily calculate the spectral function
\begin{equation}
A(\bk ,\omega )=-{1\over\pi}{\rm Im} G(\bk ,\omega ).
\end{equation}

A detailed numerical study of Eqs.(32)--(33) has been carried out in
Refs.64, 66 and 67 with the following main conclusions: There is a
well--defined quasi--particle propagating on an AF background. The
spectral function shows several coherent peaks plus some incoherent
states at higher energies. The separation between peaks as well as the band
width scales as $(J/t)^{2/3}$ for both Ising and Heisenberg cases. This has
been checked on clusters of different size ($4\times 4, 8\times 8,
16\times 16$, {\it etc}.) to make sure that the above features are not due to
finite size effects. The results for small $J/t$ are in good agreement with
those of exact diagonalization on small clusters. A similar approach using
$1/z$ expansion with $z$ as the coordination number has also been developed
\cite{67} with comparable results.

There were several approximations involved in this approach: linear spin
wave expansion and the neglect of crossing diagrams. The former does not
guarantee the hard--core boson nature of the spin excitations, which , in
turn, may lead to violation of the single occupancy constraint or
appearance of extra
degrees of freedom. As discussed in Ref.64, this will affect the short
distance behaviour (as we will see later in Section IV), but may not change
the properties for small $\bk$ near the band bottom, especially for the
Heisenberg case. The single occupancy constraint has been implemented in the
variational calculation of Sachdev \cite{68}. As for crossing diagrams, it
has been shown in Ref.67 that vertex corrections due to two-loop diagrams
vanish, while the effects of three--loop diagrams are small as found from
numerical calculations. On the other hand, the distortion of the spin
background in the presence of the hole, neglected in this approach, may have
important consequences, as we will see later, in Section IV.
\vskip0.5truecm

\noindent 3.4 Numerical approaches

Apart from the analytic approaches discussed above, there have been
extensive numerical studies on the single hole problem, using mainly two
types of techniques: exact diagonalization of small clusters and quantum
Monte Carlo simulations.

The exact diagonalization using the Lanczos--type algorithms has been
carried out by different groups for clusters of 8,10,16,18 sites
[41, 70--76]. Recently, the number of sites has been increased
up to 26 \cite{76},
for which the maximum power of the current generation of supercomputers
has been fully exploited. (The dimension of the Hilbert space is
of the order of $10^7$ and
the calculation requires at least 6 G bites disc space.) The advantage of the
exact diagonalization is the absence of the ``fermion minus sign problem''
inherent to the quantum Monte Carlo technique. On the other hand, the
limitation to small clusters may give rise to artifacts due to finite sizes.

Now the results seem to converge: There are well defined quasi--particles
with energy minima at $(\pm {\pi\over 2},\pm {\pi\over 2})$ on the
boundary of the reduced Brillouin zone. The effective mass is strongly
anisotropic: small along the zone boundary and big in the perpendicular
direction. Apart from the ground state there are some excited states for
certain momenta as seen from the spectral function. The higher excited
states have intrinsic width due to the probability of decaying into lower
states. We will discuss later how to interpret these levels. A finite size
scaling study has been performed for clusters up to 26 sizes \cite{76} (see
also \cite{77}), and the wave function renormalization factor $Z$ seems to
remain finite in the thermodynamic limit. The scaling properties of excited
states are less conclusive, especially near the zone boundary.

Another related approach is to use a truncated Hilbert space \cite{55}
which allows one to study bigger clusters using less computing power. This
technique has also been used in Refs.79,80 to study the hole bands and the
spin distortions around the hole. In general, the results obtained using this
technique are consistent with those of exact diagonalization and the
self--consistent Born approximation, considered in the previous section.

The quantum Monte Carlo technique has not been used extensively for
studying the single hole issue mainly because of the fermion sign problem.
Sorella \cite{78} has, however, systematically studied this issue using the
so--called projection Monte Carlo technique \cite{81}. He has proved a
rigorous upper bond $Z\leq 1/2$ for the $t-J$ model and has generalized the
Migdal theorem relating $Z$ to the jump of the momentum distribution at
the Fermi surface $\Delta n_k=Z$ for the 2D $t-J$ model. Using this
relation and choosing appropriate trial wave functions, he succeeded to
minimize the sign problem and to carry out a finite size scaling study for
$Z$. It seems that $Z$ vanishes in the thermodynamics limit, to the
contrary of the results from a similar study using the exact
diagonalization, just described above. Whether this discrepancy is due to
limitations of one of these techniques, is an open question for further
studies.

The Green's function Monte Carlo technique has also been used to study
the single hole problem on a $4\times 4$ cluster \cite{82}. The energy
seems to agree well with the exact result. Another alternative is to use the
Gutzwiller--projected trial wave function to perform the variational Monte
Carlo calculation \cite{83}. The location of the energy minimum and the
band width thus obtained is consistent with the known results.
\vskip2truecm

Before concluding this overview section, we would like to briefly mention
the alternative itinerant approach.
Starting from the Hubbard model and using the
Hartree--Fock approximation, one obtains the spin--density--wave (SDW)
ground state for half--filling. The one loop correction (random phase
approximation) then gives rise to the collective excitations (spin waves). This
approach has been used by Schrieffer, Wen and Zhang \cite{84} to propose the
``spin bag'' model for high--$T_C$ superconductivity. These ``spin--bags''
are nothing but spin polarons we are considering in this paper. The spin
wave velocity they have obtained is the same as what follows from the
strong--coupling $t-J$ model. This approach has been pursued by many
authors \cite{85}. The question on where the energy minimum is located, at
$(\pm {\pi\over 2},\pm {\pi\over 2})$ or $(\pm\pi ,0), (0,\pm\pi )$, depends
on which spin fluctuations (longitudinal or transverse) are considered. It is
interesting to point out that the hole--hole correlations calculated in this
approach (see paper of Frenkel and Hanke in Ref.86) are the same as follows
from Shraiman--Siggia's approach \cite{57}.

\section{QUANTUM BOGOLIUBOV--DE GENNES FORMALISM FOR THE SPIN
POLARON}

So far we have discussed the background of the single hole problem and
have overviewed various approaches
in studying this problem. In this section we
outline the quantum BdeG formalism developed by
us for this purpose, summarize the main results obtained and discuss the
open issues.
\vskip0.5truecm

\noindent 4.1 Questions to be answered

Basically, there are two sets of questions to be answered: (1) How will the
holes affect the spin background? What is the local and long--distance
distortion around the hole? How would the spin excitations and spin
correlations be modified in the presence of a dynamical hole? (2) What are
the properties of the holes (energy spectrum, effective mass, etc.) which
are embedded in this background? In view of the $\delta\to 0$ limit, one
might think of foregoing the first set of questions, assuming the ground
state of a half--filled system. However, this is not completely true,
because, as we will see later, the hole states and the distortion of the
background should be considered in a self--consistent way, which has not
been materialized in analytical formalisms reviewed in the previous section.

In fact, there are three major effects to be considered for the hole motion,
namely, (1) the quantum fluctuations of the spin system; (2) the
renormalization effect due to virtual process of emitting and reabsorbing
the spin excitations; (3) the local distortion of the spin background. Factors
(1)
 and (2) have been considered in the self--consistent Born
approximation (Section 3.3), but not
factor (3). The quantum BdeG formalism we proposed \cite{86} treats all
these three factors on an equal footing and is aimed to answer both sets of
questions. Of course, the treatment is not exact and we will discuss later
its limitations and open problems.
\vskip0.5truecm

\noindent 4.2 Lattice relaxation approach and its generalization

We have mentioned in Section 3.3 the analogy with the lattice polarons.
It is interesting to compare in more detail the properties  of
these two kinds of polarons.
For the convenience of reference we write down
here explicitly the model Hamiltonian for small lattice polarons \cite{87}:
\begin{equation}
H=H_e+H_{ph}+H_{e-ph},
\end{equation}
\[ H_e=-t_0\sum_{\langle i,j\rangle}\ C^\dagger_i\ C_j\quad,
H_{ph}=\sum_q\ \hbar\omega_q\big( b^\dagger_qb_q+{1\over 2}\big), \]
\[ H_{e-ph}=\sum_l\ C^\dagger_l\ C_l\ \sum_q\ M_q\
e^{i\bq\cdot{\bl}}(b_q+b^\dagger_{-q}), \]
where $M_q$ is the electron--phonon coupling, $\omega_q$ is the
longitudinal optical phonon frequency, while the rest of notation is
standard. In the anti--adiabatic limit $\hbar\omega_q\gg t_0$, one can use
the Lang--Firsov transformation \cite{88}
\begin{eqnarray}
H & \to & e^{-{\hat U}}\ He^{\hat U} , \\
& & \nonumber \\
{\hat U} & = & \sum_{q,l}\ {M_q\over\hbar\omega_q}\ e^{i\bq \cdot\bl}\
C^\dagger_l\ C_l (b_q-b^\dagger_{-q}), \nonumber
\end{eqnarray}
to decouple the electron and phonon degrees of freedom to obtain an
effective Hamiltonian for the electron ``dressed'' by the phonon cloud.

Comparing (35) with our $t-J$ Hamiltonian (10) and (31), the exchange
integral $J$ is, to some extent, equivalent to the phonon energy
$\hbar\omega_q$ in the lattice polaron problem, and a large $J$ means a
stiff spin background. However, there is no ``bare'' hopping term $t_0$ in
the spin polaron case.  As discussed in Section 3.3, the hole
hopping on AF background will certainly lead to frustration of the
spin background. Therefore one cannot use  the  transformation
${\hat U}$ to decouple the spin and charge degrees of freedom even in the limit
$J\gg t$. Moreover, we are more interested in the opposite limit $J\ll t$,
corresponding to realistic parameters in oxide superconductors. One might
think that it would correspond to a weakly coupled large lattice polaron.
Unfortunately, this is not true, again due to the absence of
a bare hopping term. In
a sense, there are no ``large'' spin polarons at all.

Nevertheless, the experience learned from the study of electron--lattice
coupling is still very instructive. Let us recall the lattice relaxation theory
of multi--phonon processes, originally proposed by Huang and Rhys
\cite{89} to interpret the F--center spectra and later elaborated by many
others. The crucial ingredient of their approach is to take into account the
difference of the symmetry breaking (lattice relaxation) in the initial and
final states of electronic transitions. Two of us \cite{90} have generalized
the lattice relaxation theory to study
localized excitations like polarons and solitons in quasi--one--dimensional
systems, where the self--consistency of the electronic states with lattice
configurations is essential. It turned out that the BdeG formalism
\cite{91,92}, previously developed for studying inhomogeneous
superconductivity, is a suitable framework for this purpose. Essentially, it
is a Hartree approximation in which the electronic states are solved
self--consistently with the quasi--classical motion of the lattice.

Here, in the spin polaron problem, the spin distortion
\begin{equation}
\beta_i=\langle b_i\rangle\sim\langle b_i\exp\big(-i\int H_t\ d\tau\big)
\rangle_0
\end{equation}
caused by the hole is equivalent to the lattice relaxation. The interesting
case $t\gg J$ corresponds to, in some sense, the adiabatic limit. However,
one cannot use the Hartree approximation as in quasi--1D systems, because
$t$ is not a scale of electron free motion, but rather an interaction
constant. Therefore, one should start from the opposite end, the large $J$
limit, expand everything in terms of $t/J$ and then carry out an infinite
resummation to recover the renormalized result.

In our studies we have adopted the following strategy: First consider a fixed
hole. Assuming that the spin wave spectrum does not differ significantly
from the unperturbed case, we derive a system of self--consistent BdeG
equations to determine the spin distortion and the energy spectrum of the
hole internal motion. In doing so the effects of static spin distortion, spin
quantum fluctuations and renormalization due to virtual processes are all
taken into account. Then by recovering the translational invariance and
calculating the spin configuration overlap integral, we  find the band
structure and effective mass for a propagating hole. Finally, we use these
solutions to recalculate the spin--wave spectrum and spin correlation
functions. If the spectrum is not modified significantly by the presence of
the hole, the self--consistency of the proposed scheme is justified.
\vskip0.5truecm

\noindent 4.3 Basic equations

For the convenience of reference, we rewrite our working Hamiltonian as
$H=H_J+H_t$
\begin{eqnarray}
H_J & = & J\ \sum_{i,j}\ I_{ij}\left[ b^\dagger_ib_i+{\alpha\over
2}(b^\dagger_ib^\dagger_j+b_ib_j)\right] P_iP_j,\\
H_t & = & t\ \sum_{i,j}\ I_{ij}\ h^\dagger_j\ h_i(b^\dagger_i+b_j),
\end{eqnarray}
where $P_i=1-h^\dagger_ih_i,I_{ij}=1$ for the nearest neighbours and zero
otherwise. The presence of the projection operator $P_i$ is essential for
the calculation of the spin properties.

Let us start from the case when $t>J$. As mentioned earlier, in this case
we can use the Born--Oppenheim approximation, {\it i.e.},  decompose the
combined $n$--th state of the hole--spin system as a direct product
\begin{equation}
\vert\psi_n(\bR_0)\rangle =\vert n,\bR_0\rangle_h\otimes\vert\tilde
n,\bR_0\rangle_s,
\end{equation}
where $\vert n,\bR_0\rangle_h$, is the $n$--th eigenstate of a hole
centered at $\bR_0$, whose local representation is given by
$\varphi_n(\bi-\bR_0)$ to be determined later, while $\vert\tilde
n,\bR_0\rangle_s$ is the corresponding spin part. It is important to note
that the spin configurations are different for different $n$. For the moment
we fix $\bR_0$ at the origin.

The hole states $\vert n,\bR_0\rangle_h$ and the corresponding spin
distortions are determined by BdeG type equations. Unlike the case of
quasi--1D systems, we have to go beyond the Hartree approximation. The
generalization of this type of equations including the Fock and RPA
corrections, has been worked out earlier \cite{93}, using the closed
time--path Green's function formalism. We refer the reader to the review
article {\cite{93} for technical details. The basic tool is to use the spectral
representation and the Dyson equation for the Green's function. In particular,
the retarded Green's function for the hole is defined as
\begin{equation}
G_r(j,k;t-t')=-i\ \theta (t-t')\langle \{h_j(t),h_k^\dagger(t') \}\rangle ,
\end{equation}
with its spectral representation
\begin{equation}
G_r(j,k;\omega )=\sum_m\ {\varphi_m(j)\varphi^*_m(k)\over\omega
-E_m+i\eta},
\end{equation}
where $\varphi_m$ and $E_m$ are $m$--th eigenstate and eigenvalue of the
hole, respectively. In the Hartree--Fock approximation the Dyson equation
can be written as
\begin{eqnarray}
G_r(i,j;\omega ) & = & G^{(0)}_r(i,j;\omega ) \nonumber \\
& &\quad +\sum_{l
,l'}\ G^{(0)}_r(i,l ;\omega )[\Sigma^H_r(l,l ';\omega
)+\Sigma^F_r(l ,l';\omega )]\ G_r(l ',j;\omega ),
\end{eqnarray}
where
\begin{eqnarray}
\Sigma^H_r(i,j;\omega ) & = & t\ I_{ij}\ [\beta_i+\beta^*_j], \\
\Sigma^F_r(i,j;\omega ) & = & {t^2\over N}\ \sum_{l ,l ',q}\
e^{iq\cdot (i -j )} \times \nonumber \\
& &\qquad I_{il}\ I_{jl '}\ \sum_m\Biggl\{ [1-N(E_m)]\
{F_{ll '}(u_q, v_q)\over\omega -\omega_q-E_m+i\eta} \nonumber
\\ & &\qquad +N(E_m)\ {F_{ll '}(v_q, u_q)\over\omega
+\omega_q-E_m+i\eta}\Biggr\}\ \varphi_m(l ) \varphi^*_m(l ')
\end{eqnarray}
are the Hartree (Fig.5a) and the Fock (Fig.5b) self--energy parts,
respectively, with
\begin{equation}
\beta_i=t\sum_{j ,l, m}\ N(E_m)\ \varphi_m(l )\ \varphi^*_m(j)\
I_{l j} [g_r(i,l ;\omega =0)+g'_r(i,j;\omega =0)]
\end{equation}
as spin distortion and
\begin{equation}
F_{l,l'}(u_q, v_q)=u^2_q+u_qv_q(e^{-iq\cdot (l '-j )}
+e^{iq\cdot (l -i )})+v^2_qe^{iq\cdot (l -l ' -i +j )} .
\end{equation}
Substituting (42) into (43) and remembering that
\begin{equation}
G^{(0)}_r(j,k;\omega )={\delta_{jk}\over\omega +i\eta} ,
\end{equation}
multiplying both sides of (43) by $\omega +i\eta$ and let $\omega\to E_n$,
we obtain the following  BdeG equation
\begin{equation}
E_n\varphi_n(i)=\sum_j\
[\Sigma^H_r(i,j;E_n)+\Sigma^F_r(i,j;E_n)]\varphi_n(j)
\end{equation}
to determine the energy and wave function of the hole states. In
Eq.(46), $g_r(j,k;\omega )$ and $g'_r(j,k;\omega )$ are the Fourier
transformations of the retarded spin Green's functions, defined as
\begin{mathletters}
\begin{eqnarray}
g_r(j,k;t-t') & = & -i\ \theta (t-t') \ \langle [b_j(t),b^\dagger_k(t')]\rangle
,
\\
g'_r(j,k;t-t') & = & -i\ \theta (t-t')\ \langle [b_j(t),b_k(t')]\rangle ,
\end{eqnarray}
\end{mathletters}
while $u_k, v_k$ are given by (15) and $N(E_m)$ is the hole state
occupation (0 or 1 at zero temperature).

The spin distortion $\beta_i$ can be determined via a canonical
transformation
\begin{equation}
{\cal U}=\prod\limits_i\exp (\beta_ib^\dagger_i-\beta^*_ib_i),
\end{equation}
leading to
\begin{equation}
{\cal U}\ b_i\ {\cal U}^\dagger =b_i-\beta_i,
\quad {\cal U}\ b^\dagger_i\ {\cal U}^\dagger
=b^\dagger_i-\beta^*_i .
\end{equation}
We can define a ``distorted'' N\'eel vacuum
\begin{equation}
\vert\widetilde N\rangle = {\cal U}\vert N\rangle
\end{equation}
and new spin operators
\begin{equation}
\tilde b_i={\cal U}\ b_i\ {\cal U}^\dagger
\end{equation}
so that
\begin{equation}
\beta_i=\langle\widetilde N\vert b_i\vert\widetilde N\rangle = - \langle
N\vert\tilde b_i\vert N\rangle .
\end{equation}
The true ground state with quantum fluctuations can then be represented as
\begin{equation}
\vert\tilde 0\rangle =\prod\ u^{-1}_k\exp (-\lambda_k\ \tilde
b^\dagger_k\ \tilde b^\dagger_{-k})\vert\widetilde N\rangle
\end{equation}
with $\tilde b_k$ as Fourier transformation of $\tilde b_i$ and
$\lambda_k=-v_k/u_k$. We note that the unitary transformation ${\cal U}$ and
the Bogoliubov transformation (11) commute with each other, so all the
above operations are well defined.

Before proceeding further, a few comments regarding the BdeG equation (49)
are in order. (i) The Born--Oppenheim approximation (40) used here is
different from the standard static case, because the ``potential'' in the BdeG
equation (49) is energy dependent, including some renormalization effects
due to virtual spin excitations. Therefore, the retardation effects of
spin--hole interactions have been incorporated to some extent. In fact, this
is the reason why we can get qualitatively correct results even in the range
of $t$ comparable with $J$, when the usual adiabatic approximation would
breakdown. (ii) Strictly speaking, Eq.(49) is not an eigenvalue problem due
to the energy dependence of $\Sigma^F_r$. Nevertheless, we can find
self--consistent solutions. They are dynamical generalizations of states in
the confining ``string potential'' discussed in Section 3.1. (iii) The
Brinkman--Rice results can be easily reproduced in the limit $J/t\to 0$
within our formalism. In fact, in this limit $\varphi_n(i)$ becomes an
extended state, so the retarded Green's function turns out to be
$G_r(\omega )\ \delta_{ij}$. In that case, the Hartree contribution to the
self--energy vanishes, while the Fock part becomes $\delta_{ij}\ Zt^2\
G_r(\omega -ZJ)$. One can then find $G_r(\omega )$ and the spectral
function in the Brinkman--Rice limit.
\vskip0.5truecm

\noindent 4.4 Hole states with a fixed origin

The BdeG equation (49) has been solved by us numerically, first for
$0.6>\alpha\geq 0$ \cite{86}, then for arbitrary $1>\alpha\geq 0$
\cite{94}. Moreover, the Schwinger boson formalism has been applied for
the single hole problem in the Heisenberg
limit $\alpha =1$ \cite{95}. The results of
all these calculations confirm each other and we summarize here some
main features, which are compared with those obtained in other
approaches.

There is a series of self--trapped states emerging
as solutions of (49). They are
similar to bound states of a hydrogen atom. The lowest state is of
$s$--symmetry, while the states immediately above it  are
doubly degenerate, of $p_x$ and
$p_y$ symmetry. The higher states $d_{xy}$ and $d_{x^2-y^2}$
are not degenerate, but the
energy difference is small. The ground state wave function is rather
localized in the Ising case $\alpha =0$, and becomes more extended as
$\alpha$ increases (Fig.7). The extent of the wave function is not sensitive
to $J/t$ within the range $(0.2-2)$. The ground state energy $E_0$ and the
separation between this level and the second level $E_1$ as a function of
$J/t$ is plotted in Fig.8 for the Ising case. Upon recovering the
translational invariance these states will become propagating states to
form bands, as we will see in the next section, but let us first understand
the origin of coherent propagation even for the Ising case.

Starting from the hopping Hamiltonian (38), one can derive an equation
of motion in the Hartree approximation
\begin{equation}
E_n\ h_i={\delta \ H_t\over\delta\ h^\dagger_i}=t\ \sum_i\ I_{ij}
[\langle b_i\rangle +\langle b^\dagger_j\rangle ] h_j.
\end{equation}
Since the hole wave function is not confined to one site, the spin
distortion $\langle b_i\rangle$ will not vanish on the neighbouring sites.
So the coherent motion is due to a self--generated hopping, which is
more efficient than the ``Trugman cycle'' discussed in Section 3.1. This
effect can be clearly seen by comparing Figs.7a and Fig.11a and is
one of the major consequences coming from the explicit account of spin
distortion (the Hartree term, neglected in the self--consistent Born
approximation). On the other hand, the Hartree term alone will not do the
job. As seen from Fig.8, the ground state energy goes like $-t^2/J\sim -U$,
which is a wrong energy scale. The inclusion of the Fock term reflects
higher order virtual fluctuations which bring the energy scale back to $t$,
and eventually to $J$.

The presence of a set of self--trapped states is in agreement with the
results of exact diagonalization on small clusters [41, 70--78] and the
self--consistent Born approximation  [62--64, 66, 67]. A detailed comparison of
the $J/t$ dependence requires some further studies. However, our
formalism provides explicit information about the spin distortion. In fact,
the expectation $\beta_i$ (see Eq.(55)) is a real number in the ground
state, so $\langle S_x\rangle$ is finite, while $\langle S_y\rangle =0$.
This means that the spin configuration in the ground state is a ``canted''
state, rather than a ``spiral'' state (see Fig.9), as suggested by Shraiman and
Siggia \cite{57} and further elaborated by other groups \cite{96}. We
believe that, as far as the near--by spin distortion is concerned, our result
is
correct. In fact, it agrees with an early consideration of de Gennes \cite{48}
using the $s-d$ model for discussing the spin polaron problem. The canted
state corresponds to a ferromagnetic component on an AF background.
Therefore, to consider the spin polaron as a small region with fully
polarized spins is an over simplification. What happens in reality is the
appearance of a small ferromagnetic component favoring the kinetic energy.
A recent path integral treatment of small polarons by Auerbach and Larson
\cite{59} confirmed this result. They found a small ferromagnetic region of 5
sites with probability of hole occupancy very close to ours,
as mentioned in Section 3.2.

A dipolar distortion was found in the quasi--classical treatment
\cite{57} as a long distance behaviour. Strictly speaking, the spiral state is
irrelevant for the single hole problem, since the pitch $\sim 1/\delta$
would be infinity in this case. Whether the spiral state
is a ground state for some finite doping, and
how would the long distance distortion match with the short distance
behavior,  are  all open questions. It is, however, worthwhile to
mention that the dipolar distortion was not found in the path-integral
calculation \cite{59} and the spiral phase is unstable due to strong local
distortions not included in the mean field theory \cite{97}.
\vskip0.5truecm

\noindent 4.5 Hole states as continuum bands

So far we have discussed only hole states with a fixed origin. One can
easily form Bloch states to recover the translational invariance, like
small polarons in the lattice case. The Bloch state of a propagating hole can
be written as
\begin{equation}
\vert\psi_k\rangle =(p_kN)^{-1/2}\ \sum_n\exp(i\bk\cdot\bR_n)\
\vert\psi_0(\bR_n)\rangle ,
\end{equation}
where $\vert\psi_0(\bR_n)\rangle$ is the hole state in the local
representation, while $p_k$ is the Fourier transform of the overlap integral
$p_{n,0}=\langle\psi_0(\bR_n)\vert\psi_0(0)\rangle$. As defined by (40)
$\vert\psi_0(\bR_n)\rangle$ is a direct product of the hole part and the
spin part. The energy of a propagating hole state is then given by
\begin{equation}
E_k=\langle\psi_k\vert H\vert\psi_k\rangle = p^{-1}_k\ \sum_n \exp
(i\bk\cdot\bR_n)\ E_{n,0}\langle\tilde 0,R_n | \tilde 0,0\rangle ,
\end{equation}
where the spin overlap integral $\langle\tilde 0,R_n\vert\tilde 0,0\rangle$
can be written as $e^{-S}$ with $S$ as the Huang--Rhys factor, while
$E_{n,0}$ is the hopping matrix element between hole states localized at
$0$ and $R_n$, respectively.

The energy dispersion for the polaron band formed by the ground state is
presented in Fig.10. The energy minimum is located at $({\pi\over
2},{\pi\over 2})$, in agreement with exact diagonalization [41,71--78]
and the self--consistent Born approximation [64, 66, 67]. One can also
calculate the effective mass. The results for the Ising case $\alpha =0$ are
given in Table I. From the numerical results we find that the wave function
in this case is not sensitive to the ratio $J/t$ within the range $(0.2-2)$.
By rescaling the parameters $\beta_i\to\beta_i\ J/t$ in the overlap
integral we find that the band width $\sim t/J\exp [-a(t/J)^2]$ where $a$ is
a positive constant. We note that the band width goes to zero in both limits
$t/J\to 0$ and $t/J\to \infty$. The energy separation $\Delta$ between the
lowest level and the next one shows the same tendency, as seen from Fig.8.
The limit $t/J\to 0$ is easy to understand, because the spin background is
extremely stiff, so the hole cannot move. The opposite limit $t/J\to\infty$
is more interesting. The background is soft but the coupling is strong, so
the hole carries along a huge spin excitation cloud which gives rise to a
diverging Huang--Rhys factor and a full localization.

Let us now compare the spin and the lattice polarons. If we consider $t$ as
an analogue of the electron--phonon coupling $M_q$ in (35), then the
polaron effect  diminishes in the lattice case if $M_q/\hbar\omega_q\ll
1$, while in the spin case the polaron is
completely self-trapped. The difference
is due to the presence of a bare hopping term in the lattice case (35), as we
mentioned in Section 4.2. On the other hand, the $t/J\to\infty$ limit is
quite similar, where the polaron effect is pronounced.

The functional dependence of the band width and the energy level separation
found from the exact diagonalization and self--consistent Born
approximation $(\sim (J/t)^{2/3}$ for small $J/t$, \cite{63,65,66},[72-77] ),
is different from ours, except for Ref.\cite{70}, where the authors also
reported an exponential dependence. The reason for this discrepancy has to
be understood. One possibility is that these two functional forms are valid in
different ranges of parameters. As  for our calculations we still need to
improve the self--consistency for the propagating state. The formation of
the polaron band in terms of Bloch states (57) is well-justified, if the
localized state is determined by a Hartree type self-consistent equation.
Here, however, the Fock term is crucial, so the self-consistency should be
checked directly for a given $k$ state. This problem is under consideration
now.
\vskip0.5truecm

\noindent 4.6 Effects on the spin system

The effect of single hole motion on the spin system is of the order of $1/N$
which can be ignored in many cases. However, the presence of a hole may have an
important impact on local spin properties. We have already seen in Section
4.4 the local spin distortion, but there are other effects as well. The
influence of a static hole (without $t$--term) has been studied by several
authors \cite{98}, whereas the effects of a dynamical hole have been
considered only in some numerical simulations and in a mean field treatment
\cite{99}. Recently, the renormalization effect due to spin fluctuations on
the hole motion has been discussed using a static approximation
\cite{100}, as  a boson analogue of the well--known X--ray edge
problem \cite{101}. The authors of Ref. \cite{100}
found a zero--frequency local mode which
contributes substantially to the overlap integral $e^{-\lambda}$ between
the state with a hole and the ground state without it. However, the
resulting normalization factor $\lambda =0.2$ is quite modest.

The quantum BdeG formalism developed by us to study the hole motion is a
suitable tool to treat this problem. In fact, the hole motion and the
spin background are two sides  of the same
coin. The formalism  for this part was outlined
in the second paper of Ref.87, while the
results have been reported recently \cite{102}. The main task is to
calculate the magnon propagator renormalized by the hole motion. In the
lowest approximation the polarization operator $\prod$ is given by the
one--loop correction due to the hole (Fig.5c). Using the spectral
representation for the hole propagator, $\prod$ can be expressed explicitly
in terms of the solutions of the BdeG equation, described in Section 4.3.
Omitting all technical details, we summarize here the main results
\cite{94,102}.

Firstly, we have calculated the spin distortion. For the Ising case $\alpha
=0$ the distortions are restricted to the nearest neighbours of the hole
(Fig.11a). As $\alpha$ increases, they extend over a larger region (Fig.11b).
It is important that they have the same $s$--symmetry as the hole state
itself. As we mentioned earlier, this corresponds to a canted state with
real parameters $\beta_i$. We have tried other possibilities like complex
$\beta_i$, but the canted state has the lowest energy.

Then, we have calculated the spin excitation spectrum in the presence of a
hole. In all cases when $\alpha\not= 1$, we have found several local modes.
They can be classified according to symmetry as well. Again, one of the
$s$--modes
has the lowest energy. As $\alpha$ increases, some of the local modes
merge with  the spin--wave continuum, becoming resonance modes.

Furthermore, we have calculated various spin correlation functions around the
hole. Bulut {\em et al.} \cite{98} have found an amazing result for the
Heisenberg model when the AF fluctuations are reduced near the hole in the
static case. This is not true in the dynamical case. In fact, a confining
structure is formed, {\it i.e.}, the staggered magnetization is smaller near
the
hole than that far away from it, as conjectured in the ``spin bag'' model by
Schrieffer {\em et al.} \cite{84}.
Even more interesting is the induced ferromagnetic
$\langle S^x_i\ S^x_j\rangle$ correlations for the Ising case, which shows
clearly the quantum nature of these fluctuations. Similarly, the AF
correlations near the hole in the Heisenberg case is reduced, again due to
this ferromagnetic tendency. All the above results agree with those of
small cluster calculations \cite{72}.

Finally, we have used the corrected spin excitation spectrum to solve the BdeG
equation for the hole states and have found no significant difference, which
shows the self--consistency of our procedure.

\section{CONCLUDING REMARKS}

\vskip1truecm
We have made a short excursion to various theoretical
approaches exploring the single hole problem in a quantum AF. Let us now
briefly summarize what has been established and what are the open
questions for further studies.

\vskip1truecm
\noindent 1) The first question was whether a hole can propagate in a
quantum AF. The answer coming from the Brinkman--Rice treatment and
Shraiman--Siggia's quasi--classical approach was ``no'', {\it i.e.}, the wave
function renormalization factor $Z=0$. On the contrary, the self--consistent
Born approximation and our quantum BdeG approach, as well as extensive
numerical studies on clusters have shown a coherent propagation. What is
the origin of this discrepancy?
The problem has become especially acute in view of the
controversy regarding the Fermi liquid $vs$ non--Fermi liquid behaviour of
charge carriers in high--$T_c$ superconductors. In particular, Anderson has
used Shraiman--Siggia's dipolar distortion of the spin background to argue
the existence of an ``infrared catastrophe'', leading to nonrenormalizability
of the wave function and $Z=0$ \cite{103}.
It seems that the main origin of discrepancy is due
to neglecting quantum fluctuations in the quasi classical approach. In fact,
if these fluctuations are included in the retraceable path approximation,
the ``catastrophe'' disappears \cite{104}. There are still long distance
dipolar distortions, but the normalization integral converges.

There is also some difference between the numerical results of exact
diagonalization and quantum Monte Carlo simulations, as mentioned in
Section 3.4. In view of the ``minus sign'' problem with Monte Carlo
simulations and possible finite size effects in cluster calculations,
this issue
has to be further clarified.  However,  from a general physical point of
 view, it is
hard to imagine that charge excitations cannot propagate in a magnetically
long--range ordered medium.  In all known examples,
either the quantum fluctuations are
strong so as to  destroy the long range order, or they are sufficiently
weak to allow long range order and well-defined quasi-particles\cite{105}.

However, it should be very clear that the free propagation of a single hole
does not imply a  Fermi liquid behaviour for a system with finite hole
concentration. In fact, the collective behaviour of charge carriers is quite
different from that of individual particles. For example, according to the
single hole calculation, the Fermi surface would be small pockets around
$(\pm {\pi\over 2},\pm{\pi\over 2})$. However, this is not true. A
two--hole calculation in a small cluster, corresponding to 10\% doping
already exhibits a large electronic Fermi surface, compatible with the
Luttinger theorem \cite{106}. This implies that the hole--hole correlation
and the decrease of the correlation length upon doping combine to make the
single hole problem only of marginal relevance in the doping range where
superconductivity occurs.

\vskip1truecm
\noindent 2) The existence of AF  long--range  order is the main factor
leading to self--trapping (spin polaron), which can propagate coherently.
The original suggestions of fully polarized
ferromagnetic micro-region around the hole was a simplification of the
real picture. It seems to us that a local distortion of ferromagnetic nature
(canted state) is a  quite good description. The size of this distortion
corresponds,  roughly, to the size of the ``confining string''. Originally
people
thought that the behaviour of Ising and Heisenberg AF should be very different,
since the spin distortion can be cured by the spin--flip process allowed for
the
latter. It turned out, however, that the ``string'' picture is valid to some
extent
even in the Heisenberg case, at least for small $J/t$, as we mentioned in
Section
3.3. One possible explanation could be that the reaction of the spin system
$(\sim 1/J)$ is rather slow, so the string picture holds at a short time
scale.

However, there is only one parameter -- the string length in the string picture
which leads to a one--dimensional Airy equation. In the BdeG approach we
have adopted, the self--trapping is taking place in 2D and we can classify
the hole and the corresponding spin states by symmetry which should be
helpful to gain more insight.

\vskip1truecm
\noindent 3) The spin polaron and the lattice polaron look rather similar,
but there is a major difference. There exist three energy scales in the
lattice polaron problem: the bare electron hopping, the phonon energy and
the electron--phonon coupling strength, whereas in the spin problem these
are only two: the magnon energy $J$ and the coupling strength $t$. This is
the basic reason why the spin polaron problem is more complicated: Some
kind of renormalization procedure (infinite resummation of diagrams,
summation of different paths in real space, {\it etc}.) is needed to generate
the
hole hopping. Then, what is the correct energy scale for this hopping?
$J, t$, or a
new scale like $t\exp [-a(t/J)^2]$ as we speculated? This is an open question.
Probably, the answer depends on the parameter range.
This feature of the spin polaron is the main origin of difficulty in analytical
treatments. The self--consistent Born approximation treats well the
renormalization effect of a propagating state, but does not include the
local distortion, which is essential over a quite large range of $t/J$. On the
other hand, the BdeG formalism we are using is very good to guarantee the
self--consistency of a localized state with spin distortion. However, the
procedure is not fully self--consistent for a propagating state with fixed $k$.
How to take  advantages of these two approaches is a challenge for further
research.

As a general remark we would point out that the research on the spin
polaron problem is still at a beginning stage compared with its lattice
couterpart.

Finally,  we should mention that we have not considered in this article the
implications for physical observable quantities, like
optical conductivity, Raman spectra, photo--electron emission and so on.
Some of these properties have been discussed in papers quoted (see, {\it e.g.}
\cite{41}, [71--73], \cite{106}), but to summarize all the related research
work is beyond the scope of this paper. Also,  we have not touched at all
on the hole--hole correlations and possible occurrence of superconducting
fluctuations. There exists a vast literature on this issue \cite{107}, which we
have to leave  out to keep a reasonable size of the article.

\acknowledgments

We would like to thank our collaborators W.Y. Lai, C.Q. Wu and Q.F. Zhong.
Some results of our joint research have been summarized in this article. We
are grateful to many colleagues for helpful discussions shaping up our
current understanding of this issue. Among them we would especially
mention P.A. Lee and E. Tosatti.  Y. M. L.  acknowledges  financial  support
from SERC of the United Kingdom under grant No. GR/E/79798.  This work is also
supported  from  MUST/British Council under grant No. Rom/889/92/47.

\newpage

\begin{center}
TABLE I. The dependence of hole effective mass on $J/t$ for $\alpha=0$
\end{center}
\smallskip

\begin{tabular}{rrrrrrrrrrr} \hline \hline
$J/t$ & 0.1   & 0.2 & 0.3 & 0.4 & 0.5 & 0.7 & 0.9 & 1.1 & 1.3 & 1.5  \\ \hline
$m^*/m_0$ & 330 & 11.8 & 8.7 & 6.3 & 4.9 & 4.1 & 3.9 & 4.1 & 4.4 & 4.7 \\
\hline \hline
\end{tabular}

\newpage

\begin{center}
FIGURE CAPTIONS
\end{center}
\begin{description}
\item[Fig.1.] The generic phase diagram of oxide superconductors.
\item[Fig.2.] A ``mismatched'' spin string is
created when a hole moves from one site (a) to  another site (b).
\item[Fig.3.] The  Trugman loop (see Ref. 56).
\item[Fig.4.] Schematic diagram of the dipolar spin configurations (see Ref.
58).
\item[Fig.5.] Self-energy diagrams for a hole (a)  Hartree  (b) Fock terms and
    (c) the polarization diagram for the spin propagator.
\item[Fig.6.] An example of crossing diagrams.
\item[Fig.7.] Hole ground state wave functions for (a) $\alpha=0$,
(b) $\alpha=0.5$ and (c) $\alpha=0.9$ at $t/J=0.1$.
\item[Fig.8.] The ground state energy $E_0$ and its separation $\Delta$ from
the second level as functions of $ln(1+J/t)$ in units of $t$ for the Ising
case.
They are extrapolated to the Brinkman-Rice limit for very small $J/t$.  The
dot-dashed curve is the ground state energy $E_0$ with Hartree term only.
\item[Fig.9.] Schematic diagrams for (a) canted and (b) spiral states.
\item[Fig.10.] The  hole energy band (in units of $t$) corresponding  to the
lowest
bound state for
   $t/J=0.1$ in the Ising case.
\item[Fig.11.] Spin distortions around the hole for (a) $\alpha=0.0$, $J/t=0.3$
and
  (b) $\alpha=0.4$, $J/t=0.1$.  In view of the s-symmetry only one-quarter
around the
hole is shown.
\end{description}

\end{document}